\documentclass[12pt,preprint]{aastex}

\usepackage{longtable,natbib}
\usepackage[bookmarks=false]{hyperref}

\def\arcsec{{\mbox{$^{\prime \prime}$}}}

\def\S{{\sl SWIFT}}

\begin{document}
\title{Merging and Clustering of the $\S$ BAT AGN Sample}

\author{Michael Koss\altaffilmark{1,2}, Richard Mushotzky\altaffilmark{1}, Sylvain Veilleux\altaffilmark{1}, and Lisa Winter\altaffilmark{3}}
\email{mkoss@astro.umd.edu}
\altaffiltext{1}{Astronomy Department, University of Maryland, College Park, MD, USA}
\altaffiltext{2}{Astrophysics Science Division, NASA Goddard Space Flight Center, Greenbelt, MD, USA}
\altaffiltext{3}{Center for Astrophysics and Space Astronomy, University of Colorado, Boulder, CO, USA}

\begin{abstract}

We discuss the merger rate, close galaxy environment, and clustering on scales up to a Mpc of the $\S$ BAT hard X-ray sample of nearby ($z<$0.05), moderate-luminosity active galactic nuclei (AGN). We find a higher incidence of galaxies with signs of disruption compared to a matched control sample (18\% versus 1\%) and of close pairs within 30~kpc (24\% versus 1\%).  We also find a larger fraction with companions compared to normal galaxies and optical emission line selected AGN at scales up to 250 kpc. We hypothesize that these merging AGN may not be identified using optical emission line diagnostics because of optical extinction and dilution by star formation.  In support of this hypothesis, in merging systems we find a higher hard X-ray to [O \textsc{III}] flux ratio, as well as emission line diagnostics characteristic of composite or star-forming galaxies, and a larger IRAS 60 $\mu$m to stellar mass ratio.  

\end{abstract}


\section{Introduction}
	Simulations of the growth of black holes suggest that mergers of galaxies trigger the AGN phenomenon \citep{DiMatteo:2005p5934}.  Tidal torques produced during the galaxy interaction send gas into the nuclear region to feed the black hole and enhance AGN activity \citep{Domingue:2005p5492}.  Later in the merger phase the two supermassive black holes coalesce and a rapid accretion phase is entered with a burst of star formation before settling into a relaxed state.
	
	The observational evidence for mergers driving AGN activity has been contradictory and seems to depend on the luminosity of the AGN.  Clear evidence for higher incidence of mergers is seen among QSOs (Serber et al.~2006, Veilleux et al.~2009a).  Early studies of the environment of Seyfert galaxies also appeared to show an excess of close companions \citep{Petrosian:1982A}, but recent studies of typical AGN have found no evidence for higher rates of mergers or close companions \citep{Miller:2003p5537}.  For instance, X-ray studies of AGN at intermediate redshifts did not find increased levels of mergers or close neighbors \citep{Grogin:2005p5541}.   Host galaxies of AGN in the COSMOS survey do not have greater numbers of nearest neighbor galaxies or disturbed morphologies compared to normal galaxies \citep{Gabor:2009p730}.   Finally, \citet{Li:2006p5063} analyzed 90,000 local ($z<$0.1) optically selected narrow-line AGN from the Sloan Digital Sky Survey (SDSS) and found that only 1 in 100 AGN has an extra neighbor within 70 kpc when compared to a control sample.  At larger scales between 100 kpc and a Mpc, AGN were clustered more weakly than normal galaxies.
	
	The $\S$ BAT all sky hard X-ray sample of AGN is uniquely suited to test whether local AGN are found in mergers or with close companions 
that may be driving their AGN activity because it is conducted in the 14--195 keV energy band.  This band is optically thin to much of the dust and gas obscuring the AGN and thus does not suffer from many of the biases of optical emission line classification of AGN.
		
	The BAT survey has identified 461 objects of which 262 are AGN \citep{Tueller:2009p1574}.  Most of the AGN are quite close with a median redshift of $\approx$0.03.  The 22 month BAT survey has a sensitivity of $\approx2.2\times10^{-11}$ erg cm$^{-2}$ s$^{-1}$, about 10 times more sensitive than the previous all-sky hard X-ray survey (HEAO 1 A-4: Levine et al. 1984). Higher angular resolution X-ray data for every source from either the \S~X-Ray Telescope or archival data have allowed associations to be made with known counterparts in other wavelength bands for over 90$\%$ of the BAT detections.  About 15$\%$ of the AGN are newly discovered, having never been detected as AGN before at other wavelengths.  Recent studies of the sample have indicated that it may have increased rates of mergers.  For instance, \citet{Schawinski:2009p1181} found an excess of residuals in the images after galaxy model subtraction with GALFIT for 16 BAT AGN.  Winter et al.~(2009) also found 33\% of BAT galaxies as peculiar or disturbed galaxies based on visual inspection of the 9 month survey.
	
	This Letter revisits the issue of mergers driving AGN activity using an expanded sample of 181 BAT-detected AGN.  Section 2 describes our imaging and spectroscopic data and the analysis technique for measuring the incidence of nearby companions, Section 3 describes our results and whether merging galaxies may have higher levels of optical extinction or dilution by star formation, and finally the results are summarized in Section 4.

\section{Data and Analysis}
	For our analysis, we considered three samples.  We studied a sample of BAT-detected AGN galaxies, a control sample of inactive galaxies from the SDSS matched to the BAT sample, and finally a sample of type 2 Seyferts from the SDSS matched to the BAT sample.  We will henceforth refer to the three samples as BAT AGN, the control sample, and SDSS AGN, respectively.  Our BAT AGN sample consists of nearby ($z<0.05$) AGN included in the 9 and 22 month catalogs \citep{Tueller:2009p1574}, as well as newly detected sources in the 58 month catalog (Baumgartner et al.~2010, in preparation).  The total sample includes 181 BAT-detected AGN host galaxies, $\approx$90\% of the entire northern hemisphere AGN sample.  For our sample, we use a  combination of archived imaging data from the SDSS DR7 as well as our own imaging observations taken over 17 nights at the Kitt Peak 2.1 m  telescope in the \textit{ugriz} SDSS bands.  In this BAT AGN sample, 72/181 galaxies have spectral and imaging coverage of galaxy neighbors in the SDSS.  We used these 72 BAT AGN to compare to the other two samples.  We subtracted the AGN contribution using GALFIT \citep{Peng:2002p5550} for the broad-line AGN in the SDSS and Kitt Peak (Koss et al.~2010, in preparation) images in the BAT sample.
	
%Only 50 of 145 BAT AGN have spectra for nearby galaxies in the SDSS, so we use only these BAT AGN galaxies to compare to the control sample and SDSS AGN.
	
	We generated a control sample of inactive galaxies to compare apparent merger rates.    Recent studies have found that merger rates are strongly linked to stellar mass and star formation.  Geller et al. (2006) found increased star formation with smaller galaxy separation.  In addition, \citet{Patton:2008p5024} found that at least 90\% of all major mergers occur between galaxies which are fainter than L$^{\star}$.  Therefore, to construct our control sample we used galaxies in the SDSS that have matched stellar masses, \textit{g--r} colors (as a proxy for star formation), and redshift.  From the comparison sample we excluded broad-line AGN using the SDSS galaxy class and narrow-line AGN using the Garching catalog \citep{Kauffmann:2003p2818}.  We also limited the redshifts to $z>0.01$ because of the tendency of the automated SDSS photometry to shred bright galaxies into multiple components.  We selected 2 matched control galaxies for each of the 72 BAT AGN for a total size of 144 control galaxies.
	
	We also used a sample of emission line selected AGN in the SDSS for comparison, which we refer to as the SDSS AGN.  \citet{Winter:2010p6825} found that the majority (75\%) of a sample of 64 BAT AGN were Seyferts.  We therefore chose a sample of type 2 Seyferts from the Garching catalog using the emission line diagnostics of  \citet{Kewley:2006p1554}.  We matched each of the 72 BAT AGN to the SDSS Seyfert sample in terms of color, stellar mass, and redshift for a total of 72 SDSS AGN.  
		
	 We applied the same analysis technique to each of the three samples.  To determine stellar masses we used the software kcorrect \citep{Blanton:2007p3139} with the \textit{ugriz} photometry.  To determine the redshifts of possible companion galaxies we used the spectroscopic sample from the SDSS DR7.  Since there is a 55$\arcsec$ fiber collision limit in the SDSS, as well as apparent magnitude limits for the spectroscopic survey, we supplemented our  spectroscopic data for companions.  We added any spectroscopic data of galaxy companions publicly available through NED closer than a projected separation of 30 kpc.  In the range of 30 kpc to 1 Mpc we only used the redshifts of galaxy companions in the SDSS. Throughout this work, we adopt the following cosmological parameters to determine distances: $\Omega_m$= 0.3, $\Omega_\Lambda$= 0.7, and $H_0$ = 70 km s$^{-1}$ Mpc$^{-1}$.  We define apparent mergers as galaxies that show close physical pairs (with a real-space separation of $<$30 proper kpc) or clear signs of a disturbed morphology such as tidal tails or bridges between galaxies based on visual inspection of three-color images.
	 
	We measured the closest companion to each member of our BAT AGN, control, and SDSS AGN samples on scales up to 1 Mpc.  To decide whether the neighboring galaxy is at the same radial distance, we followed the criteria of \citet{Patton:2008p5024} and used radial velocity differences of less than 500 km s$^{-1}$ between the sample galaxy and its possible companion.  We also looked at the \textit{gri} composite image of each galaxy for signs of recent mergers such as tidal tails, binary nuclei, and disturbed morphologies.
	  
%	The SDSS spectroscopic coverage is limited by apparent magnitude, and galaxies which are several magnitudes dimmer or brighter may not be covered by the spectroscopic sample.  To investigate our rate of coverage we used the photometric catalog of the SDSS DR7 in the r band.  We looked at galaxies that have companions within 30 kpc, but no spectroscopy and are up to 4 magnitudes brighter or dimmer.  	
	
\section{Results}
	
	We find that 18\% (13/72) of the BAT AGN galaxies have disturbed morphologies consistent with a recent merger.  Another 4 BAT AGN (6\%) are in close physical pairs with separations of 20--30 kpc, where tidal effects are considerably weaker.  Finally, one additional BAT AGN (1\%) shows a single nucleus with signs of tidal tails.  The overall fraction of BAT AGN undergoing mergers is therefore 25\% (18/72).  A full listing of the BAT AGN galaxies in apparent mergers is in the top panel of Table 1.  In Figure 1, we show images of nine of these galaxies selected at random from Table 1.  In the control sample we find only 1\% in apparent mergers and for the SDSS AGN sample we find 4\%.  This small rate is consistent with that of \citet{Patton:2008p5024} who found merger rates of 2\% for normal galaxies at similar distances and cosmology and other studies that have found no differences in merger rates between optically selected AGN and normal galaxies.  
	
	We also searched for galaxy companions to BAT AGN outside of the SDSS spectroscopic sample using NED.  For the 109 BAT AGN with images and spectra obtained at Kitt Peak, we find a lower rate of 22/109 or 20\% in apparent mergers.  This lower number is expected because of the reduced number of spectra of galaxy companions in NED.  A listing of these BAT AGN in apparent mergers is in the bottom panel of Table 1.   	
	
	Application of our technique to an independent sample of INTEGRAL-selected AGN detected in the hard X-rays ($z<$ 0.05, Beckmann et~al.~2009) in the northern hemisphere finds a similar rate of 28\% (15/53) in apparent mergers with companions within 30 kpc or in disrupted systems.
	
	%Another important result is the stellar mass distribution of the apparent close companions within 30 kpc that will likely merge.  In our sample, x\% are found in systems with a mass ratio of $<$3, y\% are found in merger systems with a mass ratio between 3 and 10, and z\% are found in galaxies with mass ratios $>$10 (!Need to do still!).  To increase our sample size for this comparison, we used the z band luminosity ratio as a proxy for stellar mass ratio for the galaxies outside of the SDSS imaging footprint.  
	
	In addition, we looked for faint companions to BAT AGN in the SDSS photometric catalog with no spectroscopy.  We looked specifically at the magnitude difference between the galaxy and its possible companion.  Within 30 kpc we find no additional close companions within 2 mag of the host galaxy for the BAT AGN, but an additional 1\% for the SDSS AGN, and 2\% for the control sample.  Between 2 and 3 mag, we find 4\% for the BAT AGN sample, 3\% for control sample, and 3\% for the SDSS AGN.  These faint galaxies could be at higher redshifts, and the small percentage indicates we miss only a small number of true faint companions.
	
	We use the approach of \citet{Bell:2006p4999} for a rough estimate of the number of mergers per Gyr to assess the incidence of mergers. They estimate a typical merger timescale of 0.4 Gyr for a merger of two equal mass galaxies of radius 15 kpc that are within a distance of $<$30~kpc of each other.  Mergers of unequal masses will tend to take longer because of reduced dynamical friction, so Bell et al.'s approach provides an upper limit on the merger rate. Following this method, the merger rate per Gyr is the percentage of galaxies in apparent mergers divided by 0.4 Gyr or about 63\% per Gyr for the BAT AGN. This suggests that galaxy merging may be an important mechanism to power the AGN.
	
%and the timescale of the merger dependent on a mix of orbital parameters, dynamical friction etc (Patton et al. 2002, Lin et al. 2004, Naab et al. 2006)
		
	Next, we looked for the presence of companions on larger scales, between 30 kpc and 100 kpc. The cumulative distribution of nearest companion galaxies within 100 kpc can be found in Figure~2-left.  The mean nearest neighbor galaxy separations are 41$\pm$28~kpc, 72$\pm$18~kpc, and 61$\pm$24~kpc for the BAT AGN, control galaxies, and SDSS AGN sample respectively.   For galaxies with companions within 100 kpc, a Kolmogorov--Smirnov (K-S) test indicates a $<$5\% chance that the distribution of nearest neighbor distances for the BAT AGN are from the same parent distribution as the control galaxies or SDSS AGN.  This indicates that the BAT AGN have, on average, more and closer companions than the control or SDSS AGN galaxy on scales less than 100 kpc.  We confirm that SDSS AGN and normal galaxies have similar clustering and apparent merger rates (e.g., Li et al. 2006): a Kuiper test of the SDSS AGN and control galaxies gives an 87\% chance that both samples are taken from the same parent population.

	We also determined the fraction of galaxies with neighbors between 100 kpc and 1 Mpc.  A K-S test of the distribution of closest companions within 250 kpc yields a likelihood of $<$1\% that the distributions of BAT AGN and control galaxies or SDSS AGN are from the same parent population.  A Kuiper test indicates that there is less $<$5\% probability that the BAT AGN companion galaxy distances are from the same parent population as the control sample or SDSS AGN.  In addition, a statistically higher percentage of BAT AGN have neighbors at 100--250 kpc compared to the control or SDSS AGN (Figure~2-right).  All of these statistical tests indicate that the BAT AGN have closer companions than the control or SDSS AGN sample on scales less than 250 kpc.

	Next, we examined the optical spectra of the BAT AGN in more detail to test whether the hard X-ray method may be selecting different types of AGN compared to the optical emission line classification.  Optical fluxes were corrected for galactic extinction based on Balmer decrements and were taken from \citet{Winter:2010p6825}, \citet{Ho:1997p5216}, and the Garching catalog of reduced optical spectra.  The total sample includes 29 broad-line and 45 narrow-line BAT AGN.  We examined the distribution of the hard X-ray to [O \textsc{III}] $\lambda5007$ ratio for the non-merging and merging broad-line AGN (Fig.~3) and found that all but one of the merging broad-line AGN, NGC 3227, are in the higher X-ray to optical ratio bin.  Merging and non-merging systems have similar hard X-ray luminosity distributions so this larger X-ray to [O \textsc{III}] ratio in merging systems is attributed to an [O \textsc{III}]  deficit, possibly due to unaccounted optical extinction. 
	
	In the case of narrow-line AGN, we find that merging galaxies do not have higher hard X-ray to [O \textsc{III}]  ratios than non-merging galaxies.  The mean hard X-ray to [O \textsc{III}] ratio is about 10 times larger for the narrow-line compared to the broad-line AGN, though, so other factors may have a stronger influence on the ratio such as the amount of narrow-line region gas, the geometry of the torus, the scattering fraction, or different levels of absorption of the hard X-ray flux for these objects.  This higher hard X-ray to [O \textsc{III}]  ratio among narrow-line AGN contradicts the AGN unification model, unless the [O \textsc{III}]  flux is affected by orientation effects and/or is severely underestimated due to extinction.
		
	We also looked to see if a disproportionate fraction of merging systems are missed as AGN using optical emission line diagnostics. Using the classification scheme of \citet{Kewley:2006p1554} we find that 19\% of the non-merging BAT AGN are classified as composite or HII  region-line galaxies rather than AGN.  In galaxies undergoing a merger we find a rate of 33\%.  The lower merger rate in the SDSS AGN sample may therefore be due to the fact that the optical emission-line classification is biased against mergers.	 
	
	 Elevated star formation activity could dilute the AGN emission, causing the AGN to be missed using optical emission line classification.  We therefore investigated the \textit{IRAS} data (Moshir et al. 1992) to see if the level of star formation in the merging systems was higher than the non-merging systems (Figure 4).  The 60 $\mu$m is a useful tracer of strong bursts of recent star formation and is less affected by AGN emission.  We define the specific star formation rate as the logarithm of the ratio of 60 $\mu$m emission to stellar mass.    
	 
	 The mean of the specific star formation rate for merging systems is higher (30.59$\pm$0.42 erg s$^{-1} M_{\sun}^{-1}$) than for non-merging systems (30.34$\pm$0.57 erg s$^{-1} M_{\sun}^{-1}$).  A K-S test indicates a 2\% chance that the distribution of star formation ratios for merging and non-merging galaxies are the same.  A Kuiper test has a 10\% chance.  If we include upper limits of the \textit{IRAS} flux a Kuiper test indicates a 7\% chance.  These statistical tests indicate enhanced star formation activity in merging systems compared to non-merging systems.	
	
	To further investigate the possibility that BAT AGN are found in galaxies with higher merger rates than average, we looked at the level of star formation activity based on the \textit{IRAS} 60 $\mu$m emission in BAT AGN compared to normal galaxies in the control sample and redshift-matched SDSS AGN.   A larger fraction of the BAT AGN (61\%) are detected at 60 $\mu$m than normal galaxies in the control (14\%) and SDSS AGN (11\%).  We also find that 18\% of the BAT AGN are luminous infrared galaxies (LIRGS;$L_{IR}>10^{11} L_{\sun}$) and only 3\% of the control galaxies and 1\% SDSS AGN are LIRGs.  These results indicate that BAT AGN have elevated star formation activity relative to normal galaxies and SDSS AGN (AGN contamination to the 60 $\mu$m emission in LIRGs is negligible; Petric et al.~2010).
		
\section{Summary and Discussion}	  

	We find a larger fraction of BAT AGN with disturbed morphologies or in close physical pairs ($<$30 kpc) compared to matched control galaxies or optically selected AGN.  The high rate of apparent mergers (25\%) suggests that AGN activity and merging are critically linked for the moderate luminosity AGN in the BAT sample.  We also investigated why this merging rate is larger than in optical AGN samples. We find that merging broad-line AGN galaxies are preferentially found in galaxies with high hard X-ray to  [O \textsc{III}] $\lambda5007$ ratios.  We also find a higher specific star formation rate in merging systems in the BAT sample.  This suggests that these merging AGN may not be identified using optical emission line diagnostics because of optical extinction and dilution by star formation.  Additional support for this picture comes, for instance, from \citet{Goulding:2009p6170} who found that optical emission line classification may be missing 50\% of local AGN identified via mid-infrared spectroscopy with \textit{Spitzer}.  This also seems to be the case at slightly higher redshifts and luminosities among ULIRGs ($L_{IR}>10^{12} L_{\sun}$; Veilleux et al. 2009b).

{\it Facilities:}  \facility{Swift}, \facility{Sloan}, \facility{KPNO:2.1m}, \facility{IRAS}

%\begin{references}
%\end{references}
%\bibliographystyle{/Volumes/Documents/Apps2/astronat/apj/apj}
%\bibliography{/Volumes/Documents/Apps2/astronat/biball2}

\begin{deluxetable}{l l l l l l l |} 
\tablecolumns{6} 
\tablewidth{0pc} 
\label{f33}
\tablecaption{BAT AGN in Apparent Mergers} 
\addtolength{\tabcolsep}{-2pt}
\tabletypesize{\scriptsize}
\tablehead{
\colhead{Galaxy Name\tablenotemark{1}} & \colhead{$z$} & \colhead{$\log \frac{M^*}{M_\sun}\tablenotemark{2}$} &  \colhead{Dist (kpc)\tablenotemark{3}} & \colhead{Disruption\tablenotemark{4}} & \colhead{Companion\tablenotemark{5}} & \colhead{Notes\tablenotemark{6}}  }
\startdata
2MASX J09043699+5536025&0.037&10.4&9&X& 2MASX J09043675+5535515 &SDSS\\
ARP 151&0.021&9.6&10&X& SDSS J112535.23+542314.3&SDSS\\
KUG 1208+386&0.023&10&24&& 2MASX J12104784+3820393  &SDSS\\
MCG +06-24-008&0.026&10.2&30&& SDSS J104444.22+381032.9&SDSS\\
Mrk 0739E&0.03&10.4\tablenotemark{7}&2&X& NGC 3758&SDSS\\
Mrk 1018&0.043&9.7&...&X& ...&SDSS\\
Mrk 110&0.035&10&...&X& foregound star?, tidal tail&SDSS\\
Mrk 463E&0.05&10.6\tablenotemark{7}&4&X& Mrk 463W&SDSS\\
Mrk 477&0.038&9.9&19&X& SBS 1439+537&SDSS\\
NGC 0835&0.013&10.5&15&X& NGC 833&SDSS\\
NGC 1142&0.029&10.7&17&X& SDSS J025512.06-001032.9 &SDSS\\
NGC 5106&0.032&10.6&25&X& NGC 5100 NED01 &SDSS\\
NGC 985&0.043&10.6&2&X& NGC 0985 NED02&SDSS\\
UGC 03995&0.016&10.6&9&X& UGC 03995 NOTES01&SDSS\\
UGC 05881&0.021&10.8&24&& SDSS J104644.87+255502.1 &SDSS\\
UGC 06527 NED03&0.026&10.5&24&X& UGC 06527 NED02&SDSS\\
UGC 07064&0.025&10.2&30&& CGCG 158-011 NED01&SDSS\\
UGC 08327 NED02&0.037&10.9&35&X& UGC 08327 NED01&SDSS\\
\hline
2MASX J00253292+6821442&0.012&10.1&3&X& ...&\\
2MASX J11454045-1827149&0.033&10&14&X& LEDA 867889&\\
2MASX J17232511+3630257&0.04&10.35&21&& 2MASX J17232321+3630097&\\
ESO 490-IG026&0.025&10.7&...&X& ...&\\
FAIRALL 0272&0.022&10.3&19&X& FAIRALL 0271 &\\
IRAS 05589+2828&0.033&10.4&8&X& 2MASX J06021038+2828112&\\
M106&0.002&9.9&24&& NGC 4248&\\
MCG +04-48-002&0.014&10.2&24&& NGC 6921&\\
MCG -02-12-050&0.036&10.7&33&X& 2MASX J04381113-1047474&\\
Mrk 279&0.03&10.5&27&X& MCG +12-13-024&\\
Mrk 348&0.015&10.3&22&& 2MASX J00485285+3157309&\\
Mrk 520&0.026&10.4&...&X& ...&\\
NGC 235A&0.022&9.9&9&X& NGC 0235B&\\
NGC 2992&0.008&10.3&20&X& ARP 245N &\\
NGC 3227&0.004&10&10&X& NGC 3226&\\
NGC 3786&0.009&10&14&X& NGC 3788&\\
NGC 5506&0.006&10&17&& SDSS J141324.11-031155.8&\\
NGC 6240&0.024&11\tablenotemark{7}&0.9\tablenotemark{8}&X& ...&\\
NGC 7319&0.022&10&11&X& Stephan's Quintet&\\
NGC 7469&0.016&10.5&25&& IC 5283&\\
NGC 931&0.017&10.6&6&X& UGC 01935 NOTES01&\\
UGC 11185 NED02&0.041&10.2&24&X& UGC 11185 NED01 &\\
2MASX J04234080+0408017&0.048&10&6&X& ...&Uncertain\\
3C 111.0&0.048&10&24&& 2MASX J04181911+3801368&Uncertain
\\\enddata 
\tablenotetext{1}{The top section includes BAT AGN in apparent mergers (18/72) that was compared to the SDSS AGN and control sample.  The bottom section includes apparent mergers in the Kitt Peak sample (22/109) with spectroscopic coverage of companions only from NED.}
\tablenotetext{2}{Host galaxy stellar mass based on using \textit{ugriz} photometry and the kcorrect software of \citet{Blanton:2007p3139}.}.
\tablenotetext{3}{Distance to nearest galaxy companion.}
\tablenotetext{4}{Signs of disruption consistent with a merger.}
\tablenotetext{5}{NED name where available.}
\tablenotetext{6}{SDSS:  In the SDSS spectroscopic sample, uncertain:  Companion is within 2 mags of the \textit{J}-band filter mags of BAT AGN galaxy, but has no spectroscopic redshift.}
\tablenotetext{7}{Galaxy nuclei are too close to accurately separate galaxies for stellar mass.}
\tablenotetext{8}{Based on a recent \textit{Chandra} observation.}

\end{deluxetable}

\begin{figure} 
\centering 
\includegraphics[scale=0.7]{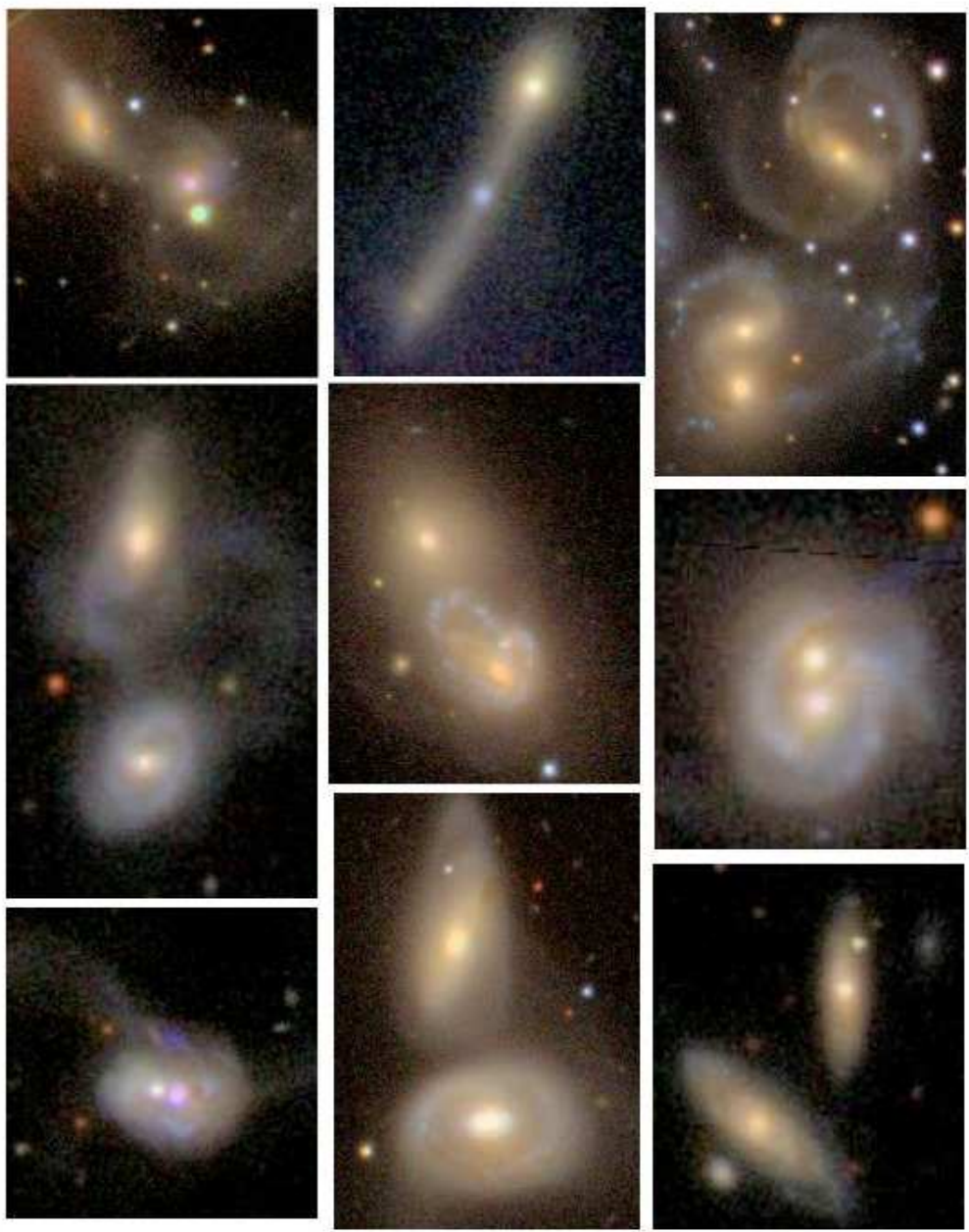} 
\caption{Composite \textit{gri} filter images of BAT AGN hosts with disturbed morphologies or companions within 30 kpc from the SDSS and Kitt Peak.  The nine galaxies were selected at random from the 40 galaxies in Table 1.    An arcsinh stretch was used as described in Lupton et al. (2004) with intensity scaled by flux.}
\label{}
\end{figure}

%Galaxies are from left to right, first column:  NGC 235A, NGC 1142, UGC 08327 NED02, Mrk 1018, ESO 490-IG026, SBS 1439+537, second column:  UGC 06527 NED03, 2MASX J09043699+5536025, NGC 985, Mrk 0739E, Fairall 0272, third column:  NGC 7319, Arp 151, 2MASX J11454045-1827149, NGC 6240, UGC 11185

\begin{figure} 
\centering 
\includegraphics[width=8.1cm]{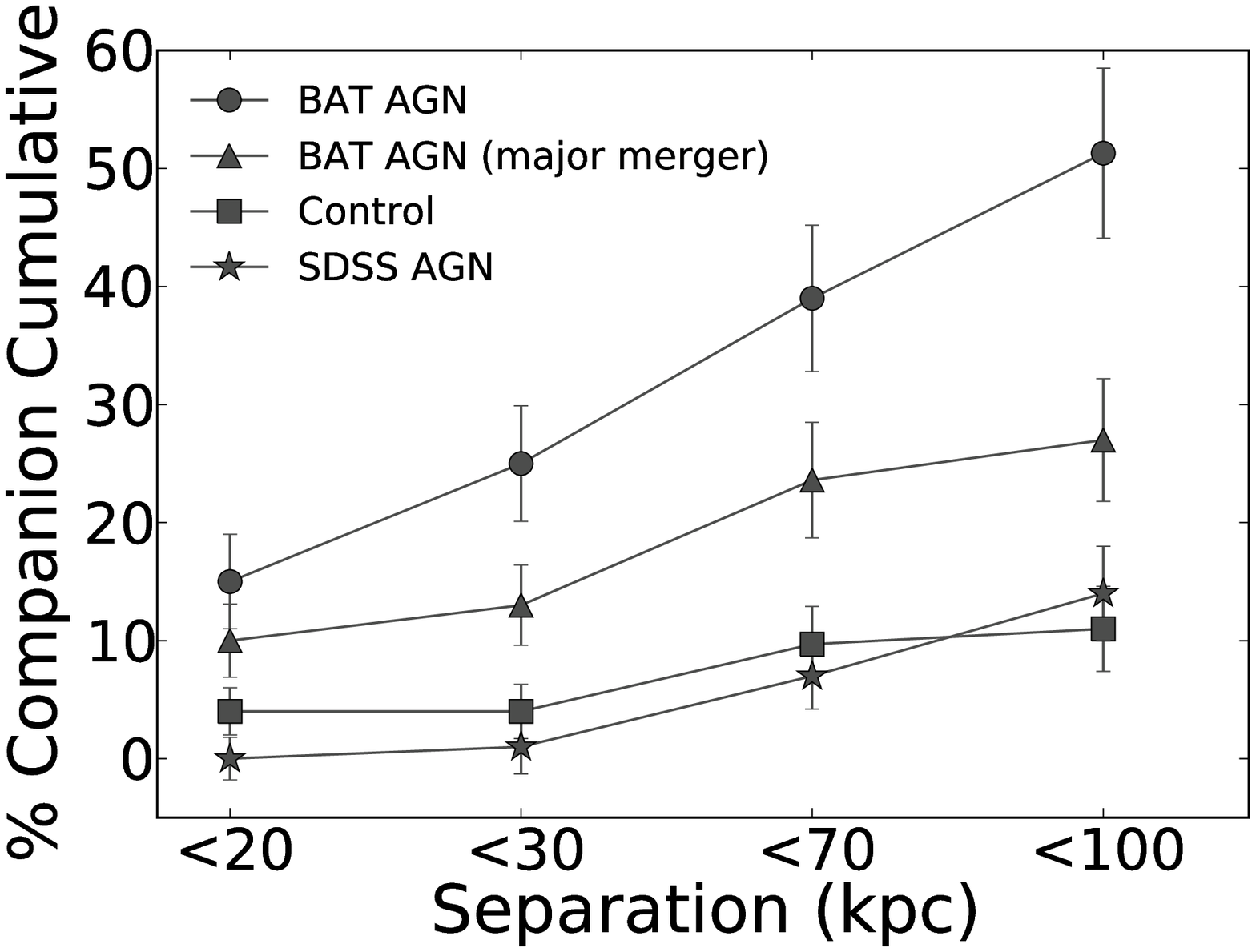} 
\includegraphics[width=8.1cm]{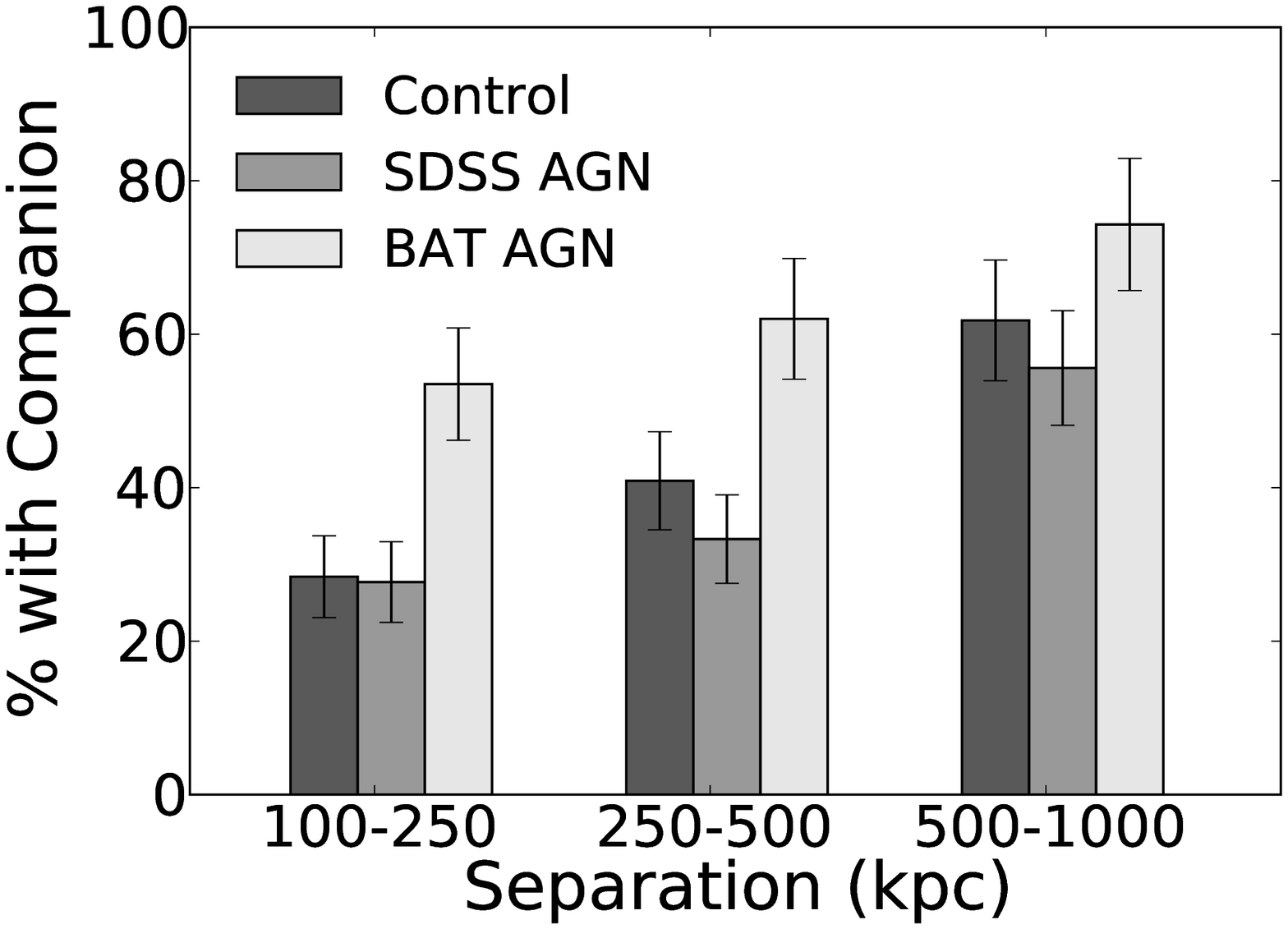} 
\caption{Left:  cumulative distributions of BAT AGN, control galaxies, and SDSS AGN with nearest neighbors identified in the SDSS survey as a function of physical separation in kpc. The error bars assume Poisson statistics. The filled circles indicate BAT AGN with any galaxy companion.  The triangle line is for AGN with a companion galaxy that has a stellar mass within a factor of 10 of the galaxy and may be considered a major merger. We find a much higher fraction of BAT AGN with close companions on scales $<$ 100 kpc. Right:  fraction of BAT AGN, control
galaxies, and SDSS AGN with companions with projected physical separations of 100--1000 kpc. An excess of companions at 100--250 kpc is seen among BAT AGN.}
\label{}
\end{figure}

\begin{figure} 
\centering
\includegraphics[scale=0.6]{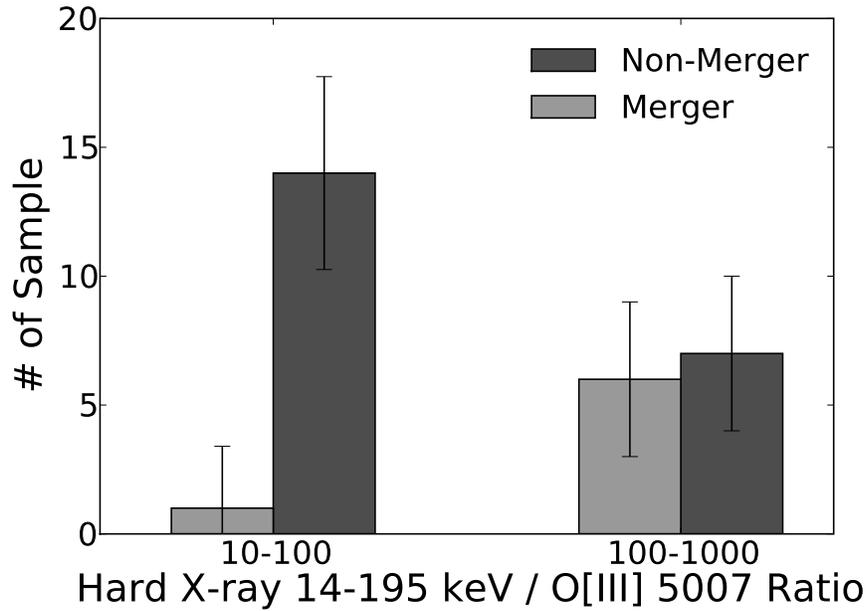} 
\caption{Histogram of the hard X-ray to [O \textsc{III}] $\lambda5007$ ratio for broad-line BAT AGN.  In order to mitigate possible systematics effects associated
with the different instrument configurations of the various surveys, we averaged the [O \textsc{III}] flux measurements from the different surveys before calculating the X-ray to [O \textsc{III}] ratios. A K-S test indicates a 3\% chance that the distributions of non-mergers and mergers are taken from the same parent population.}
\label{}
\end{figure}

\begin{figure} 
\centering
\includegraphics[width=8.1cm]{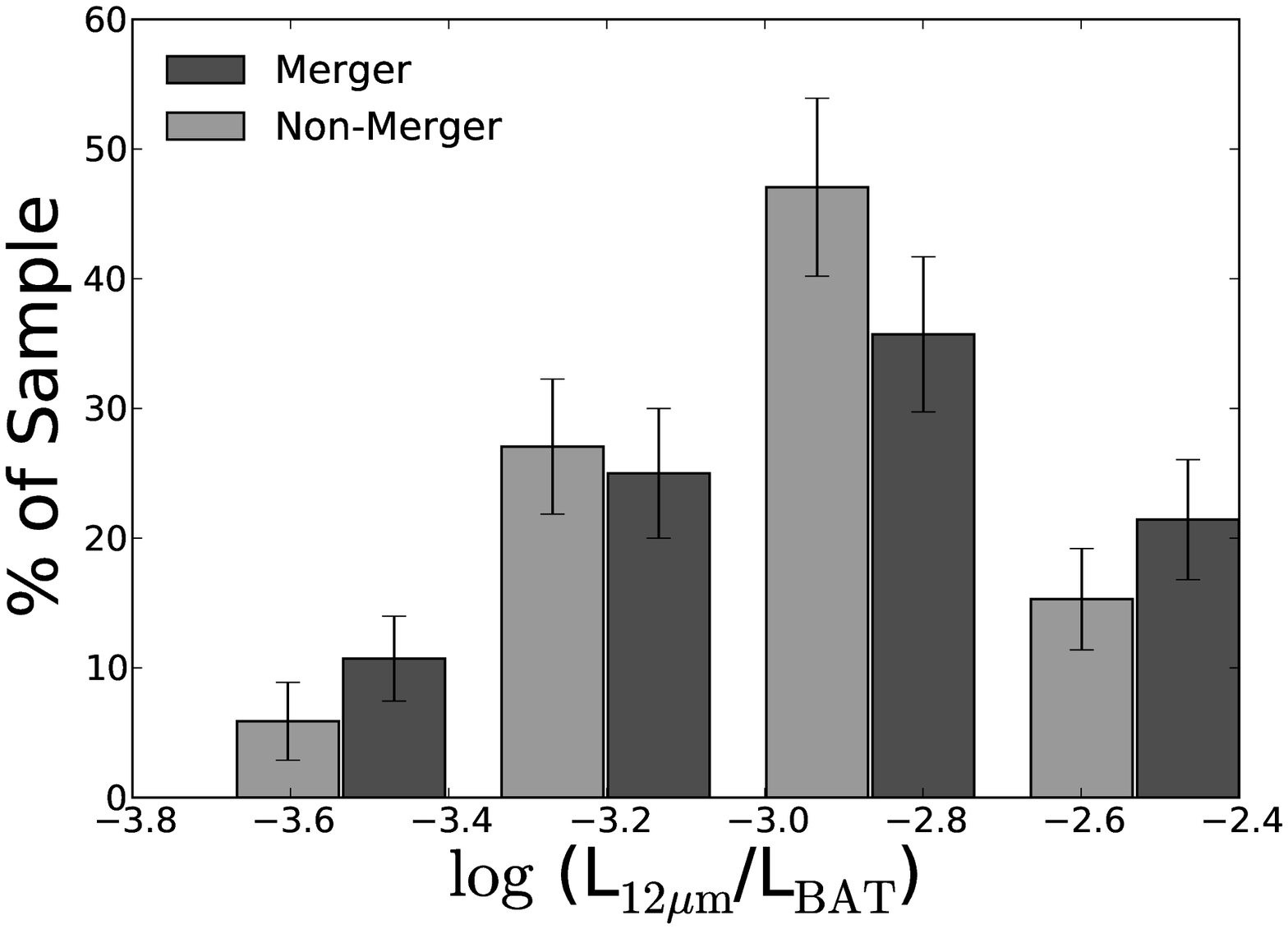} 
\includegraphics[width=8.1cm]{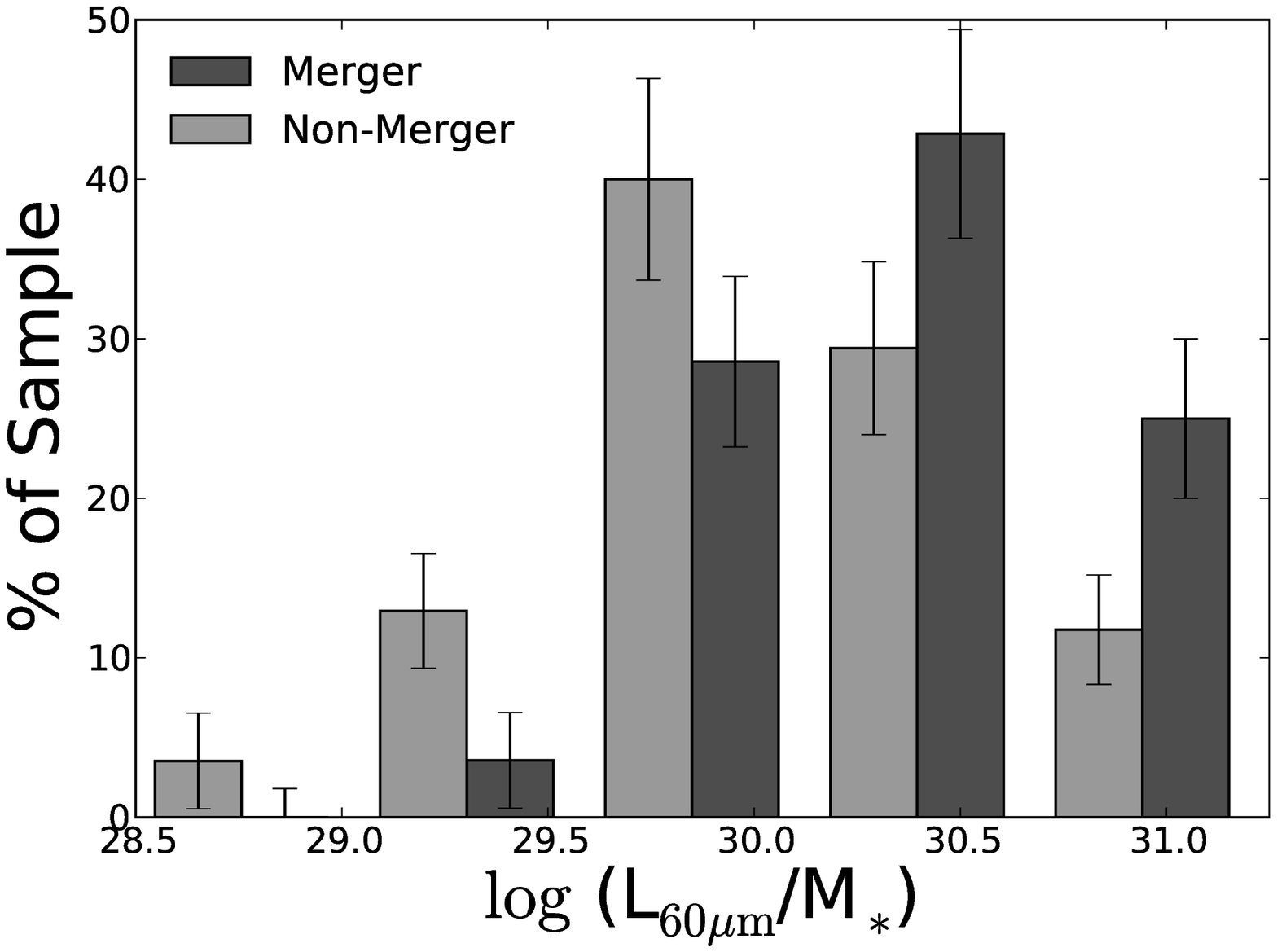} 
\caption{Left:  histogram of the IRAS 12 $\mu$m to hard X-ray emission ratio for merging and non-merging BAT AGN.  The sample includes 28 mergers and 85 non-mergers.   No difference is observed between merging and non-merging systems.  Most systems have L$_{12 \mu m}$/L$_{BAT}$ $\approx$0.1\%, confirming the known strong correlation between
the mid-IR and hard X-ray emission (e.g., Vasudevan et al. 2010).  Right:  histogram of the logarithm of the ratio of 60 $\mu$m emission to stellar mass.  This ratio is larger on average among merging systems, indicating enhanced star formation in these systems.}
\label{}
\end{figure}

\end{document}